\documentstyle[12pt]{article}
\begin{document}
\hfill HU-EP-02/56

\hfill hep-th/0212054\\
 \begin{center}
{\Large \bf
An ${\cal N}=1$ Triality by Spectrum Matching}
\vskip 0.1in
A. Misra
\footnote{e-mail:misra@physik.hu-berlin.de}\\
Humboldt-Universit\"{a}t zu Berlin\\ 
Institut f\"{u}r Physik, Newtonstra\ss e 15, 12489 Berlin (Adlershof), Germany 
\vskip 0.3 true in
\end{center}
\begin{abstract}
On promoting the type IIA side of the ${\cal N}=1$ Heterotic/type IIA dual pairs of 
\cite{VW} to $M$-theory on a `barely $G_2$ Manifold' of \cite{HM}, by spectrum-matching 
we show a possible triality between Heterotic on a self-mirror Calabi-Yau, 
$M$-theory on the above `barely $G_2$-Manifold' constructed from the Calabi-Yau on the
type IIA side and $F$-theory on an elliptically fibered Calabi-Yau
4-fold fibered over a trivially rationally ruled ${\bf CP}^1\times{\cal E}$
base, ${\cal E}$ being the Enriques surface. We raise an apparent puzzle on the F-theory
side, namely, the Hodge data of the 4-fold derived can not be obtained by a naive 
freely acting orbifold of $CY_3(3,243)\times T^2$ as one might have guessed on the basis
of arguments related to dualities involving string, M and (definition of) F theories.
There are some interesting properties of the antiholomorphic involution used in \cite{VW} for
constructing the type IIA orientifold and by us in constructing the
'barely $G_2$ manifold', that we also study.
\end{abstract}
\clearpage
\section{Introduction}

As ${\cal N}=1$ supersymmetry in four dimensions is of phenomenological interest, it 
is important to understand possible dualities between different ways of arriving at
the same amount of supersymmetry via suitable compactifications. In this 
regard, the results of \cite{VW,HLS,A,AW} are of
particular interest. While \cite{VW} construct such string dual pairs,
\cite{HLS,A,AW} also give ${\cal N}=1$ Heterotic/$M$-theory dual pairs.
As $M$-theory on $G_2$-holonomy manifolds gives ${\cal N}=1$ supersymmetry,
especially after explicit examples of the same (and Spin(7)) 
in \cite{J,C}, exceptional holonomy compactifications of $M$-theory becomes
quite relevant for the above purpose.
In the literature, so far, the ${\cal N}=1, D=4$ Heterotic/$M$-theory dual
pair constructions, stem one way or the other from the Heterotic on $T^3$
and $M$-theory on $K3$ $D=7$ duality \cite{HT,W}. Lets elaborate this point
a little. Assuming the `adiabatic argument' of \cite{VW}, the duality between 
heterotic on a $CY$ that is a $T^3$-fibration over a 3-fold and $M$-theory
on a $G_2$-holonomy manifold that is a $K3$-fibration over the same 3-fold
of \cite{A,AW} was obtained by fiberwise application 
of duality with a twist (to get ${\cal N}=1$ supersymmetry)
to the abovementioned $D=7$ Heterotic/$M$-theory duality. Similarly, by application of 
fiberwise duality, accompanied by a twist, with 
the 3-fold base above being a $T^3$, the heterotic/$M$-theory dual pair of
\cite{HLS} was obtained. The 7-manifold on the $M$-theory side in \cite{HLS} 
will be the motivation for the choice of the 7-manifold relevant to us.
Now, lets come to string-string duality in $D=6$: Heterotic on $T^4$ is
dual to type IIA on $K3$. By fiberwise application of duality using
a common ${\bf CP}^1$ base, and taking $K3$ to be elliptically fibered
($T^2$-fibration over ${\bf CP}^1$) and the $CY$ to be a $K3$-fibration over
${\bf CP}^1$, ${\cal N}=2$ dual pairs of heterotic on $K3\times T^2$ with
vector bundles on $K3$ embedded in $E_8\times E_8$, and type IIA on the $K3$-fibered
Calabi-Yau were constructed in \cite{KV}. From these dual pairs,
${\cal N}=1$ dual pairs were obtained in \cite{VW} by orientifolds
of the $CY$ on the type IIA side whose image on the heterotic side 
corresponded to the self-mirror $CY$ of \cite{Feretal}.
The question is what the ${\cal N}=1$ Heterotic/$M$-theory analog of the 
Heteoric/type IIA ${\cal N}=1$ dual pair of \cite{VW} is. 
As the $D=7$ Heterotic/$M$-Theory duality is related to
the $D=6$ Heterotic/String duality (as the decompactification limit - 
see \cite{W}), it is reasonable to think that there has to be such an 
${\cal N}=1$ Heterotic/$M$-theory dual pair. Additionally, it will be  interesting
to work out an example that is able to explicitly relate an ${\cal N}=1$ Heterotic
theory to M and F theories, as opposed to examples in the literature on only
${\cal N}=1$ Heteorotic/type IIA or Heterotic/M-theory or Heterotic/F-theory dual pairs
(See \cite{B} for some comments).
In Section {\bf 2}, we propose that the $M$-theory
side is given by a 7-manifold of $SU(3)\times{\bf Z}_2$-holonomy of the
type $(CY\times S^1)/g.{\cal I}$, where $g$ is a suitably defined 
freely-acting  antiholomorphic involution on the $CY$ which is precisely the
same as the one considered in \cite{VW}, $\Omega$ is the world-sheet parity
and ${\cal I}$ reflects the $S^1$. These 7-manifolds are referred to as
``barely $G_2$ manifolds'' in \cite{HM}. 
In addition, the $D=4,\ {\cal N}=1$ Heterotic/F-theory dual models constructed have
the following in common (as a consequence of applying fiberwise duality to
Heterotic on $T^2$ being dual to F-theory on $K3$). 
The Heterotic theory is compactified on a
$CY_3$ that is elliptically fibered over a 2-manifold $B_2$. The F-theory
dual of this Heterotic theory is constructed by considering 
an elliptically fibered Calabi-Yau 4-fold
$X_4$ that is elliptically fibered over a 3-manifold 
$B_3$. Additionally, the base $B_3$ is a ${\bf 
P}^1$-fibration 
over $B_2$ (the same one that figures on the heterotic side). In Section {\bf 3},
we propose that the required Calabi-Yau 4-fold on the F-theory side 
is elliptically fibered over a trivially rationally ruled base 
given by ${\bf CP}^1\times{\cal E}$, ${\cal E}$ being the Enriques surface.
We raise an apparent puzzle regarding the derived Hodge data and the one that one might
have naively guessed based on string/M/F dualities.
Section {\bf 4} has a discussion on the results obtained and outlook for future
work.

\section{M-Theory Dual}

In this section we construct the M-theory uplift of type IIA background of \cite{VW}.
Now, the specific ${\cal N}=1$ Heterotic/type IIA dual pair of \cite{VW}
that we will be considering in this letter is Heterotic on a $CY$ given by
${K3\times T^2\over{\bf Z}_2^E}$ and type IIA on orientifolds of
$CY$'s (the mirrors of which are) given as hypersurface of degree 
24 in ${\bf WCP}^4[1,1,2,8,12]$, the mirror duals to which, are given by:
\begin{eqnarray}
\label{eq:K3xCP1}
& &  z_1^{24}+z_2^{24}+z_3^{12}+z_4^3+z_5^2-12\alpha z_1 z_2 z_3 z_4 z_5
-2\beta z_1^6 z_2^6 z_3^6 - \gamma z_1^{12} z_2^{12}=0.\nonumber\\
& & 
\end{eqnarray}
The ${\bf Z}_2^E$ represents the Enriques involution times reflection of
the $T^2$ as considered in \cite{Feretal,HLS} given by the action
\begin{equation}
\label{eq:Z2Edieuf}
{\bf Z}_2^E: (u_1,u_2,u_3)\rightarrow(-u_1+{1\over2},u_2+{1\over2},-u_3),
\end{equation}
($u_{1,2}$ and $u_3$ are coordinates on $K3$ and $T^2$ respectively)
and the space-time orientation reversing antiholomorphic involution
used for constructing the CY orientifold is:
\begin{equation}
\label{eq:wdieuf}
\omega:(z_1,z_2,z_3,z_4,z_5)\rightarrow({\bar z}_2,-{\bar z}_1,{\bar z}_3,
{\bar z}_4,{\bar z}_5).
\end{equation}
Now, (barely) $G_2$-Manifolds of the type ${CICY\times S^1\over g.I}$
where $g$ was an antiholomorphic involution,  
were considered in \cite{PP} where $g$  was one of the following, 
the former corresponding to an involution with fixed points, leading
to $G_2$-holonomy manifolds and the latter acting freely leading to 
'barely $G_2$-manifolds":
\begin{eqnarray}
\label{eq:invdieufs}
& & (z_1,...,z_{n+1})\rightarrow({\bar z}_1,...,{\bar z}_{n+1})\ 
{\rm for}\ {\bf CP}^n,\ {\rm or}\nonumber\\
& &  (z_1,z_2...,z_{2n})\rightarrow({\bar z}_2,-{\bar z}_1,...,{\bar z}_{2n},
-{\bar z}_{2n-1})
\ {\rm for}\ {\bf CP}^{2n-1},
\end{eqnarray}
(for CICY expressed as a set of homogenous equations in a single ${\bf CP}^m$
space)
where in the second choice, it is understood that the antiholomorphic
involution acts pairwise on the homogenous coordinates. The antiholomorphic
involution that we require is $\omega$ of (\ref{eq:wdieuf}), which
is a combination of the ones in (\ref{eq:invdieufs}).\footnote{To see 
that $\omega$ is an involution, one sees that $\omega^2$ takes 
$(z_1,z_2,z_3,z_4,z_5)$ to $(-z_1,-z_2,z_3,z_4,z_5)$, which 
is projectively equal to unity as can be seen by
setting the $C^*$ variable $t$ to -1 in $z_i\sim t^{w_i}z_i$ in ${\bf WCP}^4$
homogenous coordinates.}
Another point worth keeping in mind is that under $\omega$ of 
(\ref{eq:wdieuf}), the K\"{a}hler form $J$ going over to -$J$
is only a statement in the cohomology group $H^{1,1}$. 
One can define inhomogenous  coordinates for, e.g., Y, in the $z_2\neq0$
coordinate patch:
\begin{equation}
\label{eq:coomrds}
u\equiv{z_1\over z_2};\ v\equiv{z_3\over z_2^2};\ w\equiv{z_4\over z_2^4},
\end{equation}
[using which one can solve for ${z_5\over z_2^{12}}$ from the defining
equation (\ref{eq:K3xCP1}), and hence is not included as part of the $CY$
coordinate system]. Then, one can show that 
\begin{equation}
\label{eq:transf}
J\stackrel{\omega}{\rightarrow}-J+{\cal O}\biggl({1\over|u|^{m>0}}\ {\rm or}\
g_{u{\bar u}}-{\rm independent\ terms}\biggr),
\end{equation}
such that the $-J$ and $-J+$ extra terms both belong to the same 
cohomology class of $H^{1,1}$.
As $u\in{\bf CP}^1$-base coordinate and $g_{u{\bar u}}$ gives the size of the
${\bf CP}^1$ base, in the large base-limit of \cite{VW}, $J$ under the
antiholomorphic involution $\omega$ goes over to $-J$ {\bf exactly}. Similarly,
$H^{2,1}$ goes over to $H^{1,2}$ (and $X_{2,1}\in H^{2,1}$ goes over
to $X_{1,2}\in H^{1,2}$ exactly in the large-base limit of \cite{VW})
but an element $Y^{1,1}$ of $H^{1,1}$ goes over to an element of
the cohomology class $[-Y^{1,1}]$ of $H^{1,1}$ and no statement can 
be made for large base-limit exactness like
the ones for $J$ or $\Omega$ above. 
\footnote{The exact expressions for $J$ and an element of $H^{2,1}$ under
the action of $\omega$, written in terms of inhomogenous coordinates in the
$z_2\neq0$ coordinate patch are as follows.
\begin{equation}
\label{eq:Jdieuf}
J=g_{u{\bar u}}du\wedge d{\bar u}+g_{v{\bar v}}dv\wedge d{\bar v}+
g_{w{\bar w}}dw\wedge d{\bar w}
+2i{\rm Im}\biggl[g_{u{\bar v}}du\wedge d{\bar v}+g_{u{\bar w}}du\wedge
d{\bar w}+g_{v{\bar w}}dv\wedge d{\bar w}\biggr],\nonumber\\
\end{equation}
under (\ref{eq:wdieuf}), which in terms of the $CY$ coordinates
$(u,v,w)$ is written as $\omega:(u,v,w)\rightarrow(-{1\over{\bar u}},
{{\bar v}\over{\bar u}^2},{{\bar w}\over{\bar u}^2})$, goes over to
\begin{eqnarray}
\label{eq:transfJ}
& & -J+d{\bar u}\wedge du\biggl({4|v|^2\over|u|^2}g_{v{\bar v}}
+{64|w|^2\over|u|^2}g_{w{\bar w}}
+2{\rm Re}\biggl[-{2v\over u}g_{v{\bar u}}-{8w\over u}g_{w{\bar u}}
+{16{\bar v}w\over|u|^2}g_{w{\bar v}}\biggr]\biggr)\nonumber\\
& & -4i{\rm Im}
\biggl(\biggl[{v\over u}g_{v{\bar v}}+4{w\over u}g_{w{\bar u}}\biggr]
d{\bar v}\wedge du\biggr)+2i{\rm Im}\biggl(g_{w{\bar v}}d{\bar v}\wedge
dw\biggr)\nonumber\\
& & +2i{\rm Im}\biggl(\biggl[-{8{\bar w}\over{\bar u}}g_{w{\bar w}}
+g_{w{\bar u}}-{{\bar v}\over{\bar u}}g_{w{\bar v}}\biggr]
d{\bar u}\wedge dw\biggr)
\stackrel{|u|\rightarrow\infty,{g_{u_i{\bar u}_j}\over g_{u{\bar u}}}
\rightarrow0,i,j\neq1}{\longrightarrow}-J.\nonumber\\
\end{eqnarray}
 Consider $A\in H^{2,1}(CY_3)$, written out
in components as:
\begin{equation}
\label{eq:Adieuf}
A=\sum_{i\neq j\neq k=1}^3A_{u_uu_j{\bar u}_k}du_i\wedge du_j\wedge d{\bar u}_k
+\sum_{i\neq j=1}^3A_{u_iu_j{\bar u}_i}du_i\wedge du_j\wedge d{\bar u}_i,
\nonumber\\
\end{equation}
where $u_{1,2,3}\equiv u,v,w$. Then,
\begin{eqnarray}
\label{eq:2,1transof}
& & A\stackrel{\omega}\rightarrow A^*+d{\bar u}\wedge d{\bar v}\wedge du
\biggl({2v\over u}A_{{\bar v}{\bar u}v}-{16v{\bar w}\over|u|^2}A_{{\bar v}
{\bar w}v}+{64|w|^2\over|u|^2}A_{{\bar w}v{\bar v}w}
-{8w\over u}A_{{\bar u}{\bar v}w}+{8{\bar w}\over{\bar u}}A_{{\bar v}{\bar w}u}
\biggr)\nonumber\\
& & +d{\bar u}\wedge d{\bar w}\wedge du\biggl({2|v|^2\over|u|^2}A_{{\bar v}
{\bar w}v}+{8w\over u}A_{{\bar w}{\bar u}w}-{16{\bar v}w\over|u|^2}
A_{{\bar w}{\bar v}w}-{2{\bar v}\over{\bar u}}A_{{\bar v}{\bar w}u}
+{2v\over u}A_{{\bar w}{\bar u}v}\biggr)\nonumber\\
& & +d{\bar v}\wedge d{\bar u}\wedge dv(-{8{\bar w}\over
{\bar u}}A_{{\bar v}{\bar w}v})+d{\bar w}
\wedge d{\bar u}\wedge
dw(-{2{\bar v}\over{\bar u}}A_{{\bar w}{\bar v}w})+d{\bar v}\wedge d{\bar w}
\wedge du(-{2v\over u}A_{{\bar v}{\bar w}v}+{8w\over u}A_{{\bar w}{\bar v}w})
\nonumber\\
& & +d{\bar u}\wedge d{\bar w}\wedge dv(-{2{\bar v}\over{\bar u}}A_{{\bar v}
{\bar w}v})+d{\bar u}\wedge d{\bar v}
\wedge dw(-{8{\bar w}\over{\bar u}}A_{{\bar w}{\bar v}w})
\stackrel{|u|\rightarrow\infty}
\rightarrow A^*.\nonumber\\
\end{eqnarray}}
To summarize, we get:
\begin{eqnarray}
\label{eq:omegaprops}
& & [\omega^*(J)]=[-J];\ \omega^*(J)\stackrel{\rm large\ {\bf CP}^1}{\longrightarrow}
-J,\nonumber\\
& & [\omega^*(X)]=[{\bar X}];\ \omega^*(X)\stackrel{\rm large\ {\bf CP}^1}{\longrightarrow}{\bar X};\nonumber\\
& & [\omega^*(Y)]=[-Y],
\end{eqnarray}
where $X\in H^{2,1}(CY_3\longrightarrow_{K3}{\bf CP}^1)$ and
$Y\in H^{1,1}(CY_3\longrightarrow_{K3}{\bf CP}^1)$, and $[\ ]$ denotes
the cohomology class. The closed and co-closed calibration 3-form $\phi$ is
given by:
\begin{equation}
\label{eq:caldieuf}
\phi=J\wedge dx +Re(e^{-{i\theta\over2}}\Omega),
\end{equation}
where $x$ is the $S^1$ coordinate, and $\Omega$ is the holomorphic 3-form
of the $CY_3(3,243)$. 
To get the spectrum for $M$-theory compactified on the `barely $G_2$
manifold' ${\cal Z}\equiv{CY\times S^1\over \omega.{\cal I}}$, 
one sees (See \cite{HM}) that ${1\over2}(H^{3,0}(CY)+H^{0,3}(CY))$ corresponding
to ${1\over 2}(h^{3,0}(CY)+h^{0,3}(CY))=1$, is invariant under the ${\bf Z}_2$
involution $\omega$. Similarly, ${1\over2}(H^{2,1}(CY)+H^{1,2}(CY))$ 
corresponding to 
${1\over2}(h^{1,2}(CY)+h^{2,1}(CY))=h^{2,1}(CY)$ elements, is invariant under
the involution $\omega$. As shown in \cite{VW}, $\omega$ acts as $-1$
on $H^{1,1}(CY)$ implying that $H^{1,1}_+(CY)$, i.e., the part of $H^{1,1}(CY)$
even under $\omega$ is zero, and the part odd, $H^{1,1}_-(CY)=H^{1,1}(CY)$.
Hence, $n_V$, $n_C$ that denote
respectively the number of vector and hypermultiplets, will be given by:
\begin{eqnarray}
\label{eq:vechyperct}
& & n_V({\cal Z})=h^{1,1}_+(CY)=0,\nonumber\\
& & n_C({\cal Z})=h^{2,1}(CY)+h^{3,0}(CY)+h^{1,1}_-(CY)=243+1+3=247.
\end{eqnarray}
$M$-theory on barely $G_2$ manifolds yielding $n_V=0$ have also been
considered in \cite{KKP}.
Lets now briefly review the spectrum of heterotic theory on 
${K3\times T^2\over{\bf Z}_2}$ (See \cite{VW}), 
where the ${\bf Z}_2$-involution is the Enriques involution 
on the $K3$ and a reflection on the $T^2$. Because of
the reflection on $T^2$, the four ${\cal N}=2$ vector multiplets get projected
out. The $S,T,U$ moduli  survive the involution as ${\cal N}=1$ chiral
multiplets. In addition, one gets 20 ${\cal N}=1$ chiral multiplets from
the Enriques surface moduli and 224 from the $E_8\times E_8$ gauge 
bundle moduli (assuming that interchange of the two $E_8$ factors under the
action of $\omega$)\footnote{One gets 224 ${\cal N}=2$ hypermultiplets
 by noting that the dual coexter
 for $E_8$, $C_2(E_8)=30$ and using that the total number of neutral hypermultiplets
coming from a single $E_8$ factor with $\int_{K3}c_2=12$ (for (12,12) instantons
embedded in the two $E_8$'s) is given by $30\times12-248=112$. Of each ${\cal N}=2$
hypermultiplet, one ${\cal N}=1$ chiral multiplet survives the involution.}
. Thus one gets a total of 3+20+224=247 ${\cal N}=1$
chiral multipets. 

Now coming to the spectrum of type IIA theory compactified 
on ${CY\over\omega.\Omega}$ (See \cite{VW}). Denoting the $CY$ indices by
$i,{\bar j}$ and the four-dimensional indices by $\mu$, the ${\cal N}=2$
spectrum consist of $h^{1,1}(CY)$ 
vector multiplets the scalar components of each of which
consists $(g_{i{\bar j}}, b_{i{\bar j}}, c_{\mu i{\bar j}})$ ($g_{i{\bar j}}$
from the $D=10$ metric $g_{MN}$ and $b_{i{\bar j}}$ from the 
$D=10$ antisymmetric tensor $B_{MN}$ from the NS-NS sector, and
$c_{\mu i{\bar j}}$ from the $D=10$ 3-form gauge potential $c_{MNP}$
from the RR sector), and $h^{2,1}(CY)$ hypermultiplets the bosonic
components of each of which is given by $(g_{[ij]}, g_{[{\bar i}{\bar j}]},
c_{ij{\bar k}}, c_{{\bar i}{\bar j}k})$ ($g_{[ij]}$ and its complex conjugate
from $g^{l{\bar l}}g^{k{\bar k}}X_{[i|{\bar l}{\bar k}}\Omega_{j]lk}$,
$X_{1,2}\in H^{1,2}$ and $\Omega_{3,0}$ being the nowhere vanishing
holomorphic 3-form)+ another hypermultiplet whose bosonic components
are given by $(\phi,a(\sim b_{\mu\nu}),c_{ijk},c_{{\bar i},{\bar j},{\bar
   k}})$. Among the $h^{1,1}(CY)\ {\cal N}=2$ vector multiplets, the gauge
field gets projected out under $\omega.\Omega$ 
to give $h^{1,1}(CY)\ {\cal N}=1$ chiral multiplets.
From the $h^{2,1}(CY)+1\ {\cal N}=2$ hypermultiplets, one each from the
two NS-NS and RR states survive the orientifold projection, yielding 
$h^{2,1}(CY)+1\ {\cal N}=1$ chiral multiplets. Hence,
one gets a total of $h^{1,1}(CY)+h^{2,1}(CY)+1=3+243+1=247$ ${\cal N}=1$ chiral multiplets  

This one sees that the spectra associated with Heterotic on ${K3\times T^2\over
{\bf Z}_2}$, type IIA on ${CY\over\omega.\Omega}$, and  
$M$-theory on ${CY\times S^1\over\omega.{\cal I}}$ match.

\section{F-Theory Dual}

We now show the possibility of finding an ${\cal N}=1$ triality
between the ${\cal N}=1$ heterotic on $CY_3(11,11)$(/type IIA on 
${CY_3(3,243)\over\omega.\Omega}$ dual pair) of Vafa-Witten, 
$M$ theory on the ``barely $G_2$ manifold" 
${CY_3(3,243)\times S^1\over\omega.{\cal I}}$ of $SU(3)\times{\bf Z}_2$
holonomy, and F-theory on an elliptically fibered $X_4$ , 
where the ``11,11" and ``3,243" denote the Hodge numbers, 
$\omega$ is an orienation-reversing
antiholomorphic involution, ${\cal I}$ reverses the $S^1$. The $X_4$ that we obtain
in this section will be obtained by assuming that the required F-theory dual must exist.
Of course, given the basic string/M/F dualities, we know that such an F-theory dual must
exist - what we show is given the same, what the geometric data of the required $X_4$
must be.
In principle, one should be able to get the right ${\cal N}=1$ F-theory
model by following the ${\cal N}=2$ Higgs chain of \cite{Kachruetal}:
$E_8\times E_8\rightarrow E_7\times E_7\rightarrow E_6\times E_6\rightarrow$
$SO(10)\times SO(10)\rightarrow SO(9)\times SO(9)\rightarrow SO(8)\times SO(8)$
$\rightarrow SO(7)\times SO(7)\rightarrow SU(4)\times SU(4)\rightarrow SU(3)\times SU(3)$
$\rightarrow SU(2)\times SU(2)\rightarrow SU(1)\times SU(1)$ by embedding of suitable
gauge commutants on the $K3$, or equivalently by embedding of $E_8\times E_8$ vector bundle
on $K3$ in one step, followed by tensoring with a $T^2$ and taking a 
suitable ${\bf Z}_2$-involution. One uses the Kodaira classification of singularities
to count the F-theory moduli. We however do not follow this approach in the
following.

The $CY_3$ on the heterotic side that we are interested in 
is one that is obtained by a freely-acting Enriques involution acting on the $K3$ 
times a reflection of the $T^2$, in $K3\times T^2$, i.e., the Voisin-Borcea elliptically
fibered $CY_3(11,11)\equiv {K3\times T^2\over g.{\cal I}}$, where $g$ is the generator of the
Enriques involution and ${\cal I}$ reflects the $T^2$. Hence, the $B_2$
above is ${K3\over g}$. 
Now, the ${\cal N}=2$ dual pair in \cite{KV} consisted of embedding
$SU(2)\times SU(2)$ in $E_8\times E_8$ on the Heterotic side,
resulting in $E_7\times E_7$, which is then Higgsed away. All that survives from the
$T^2$ in $K3\times T^2$ are the abelian gauge fields corresponding to  $U(1)^4$.
As shown in Vafa-Witten's paper\cite{VW}, in the ${\cal N}=1$ dual pair
obtained by suitable ${\bf Z}_2$-moddings of both sides of the ${\cal N}=2$
Heterotic/type IIA dual pair, the $U(1)^4$ gets projected out so that 
there are no vector multiplets and one gets
247 ${\cal N}=1$ chiral multiplets on the Heterotic side
on $CY_3(11,11)$. We should be able to get the same spectrum 
on the F-theory side. If $r$ denotes
the rank of the unbroken gauge group in Heterotic theory, then the number of
${\cal N}=1$ chiral multiplets in F-theory is given by the formula (\cite{M,CL}):
\begin{equation}
\label{eq:nC}
n_C={\chi(X_4)\over6}-10+h^{2,1}(X_4) - r,
\end{equation}
which excludes the $S$ modulus of the Heterotic theory. The rank $r$ in
turn is expressed as:
\begin{equation}
\label{eq:r}
r=h^{1,1}(X_4) - h^{1,1}(B_3) - 1 + h^{2,1}(B_3).
\end{equation}
For Heterotic theory on $CY_3(11,11)$, $r=0$.

The fibration structure can be summarized as:
$X_4\longrightarrow_{T^2}B_3\longrightarrow_{{\bf CP}^1}B_2\equiv{K3\over g}
\equiv{\cal E}\equiv$Enriques surface.
Given that for elliptically fibered $X_4$, 
$h^{1,1}(X_4)-h^{1,1}(B_3)-1\geq0$, $r=0$ implies that
\begin{eqnarray}
\label{eq:consts1}
& & h^{1,1}(X_4)=h^{1,1}(B_3)+1>0;\nonumber\\
& & h^{2,1}(B_3)=0.
\end{eqnarray} 

Now, in \cite{FMW}, $r=0$ was obtained by embedding an $E_8\times E_8$
vector bundle for which it was shown that the number of space-time filling 
F-theory 3 branes, needed for tadpole cacellation, matched the
number of Heterotic 5-branes needed for anomaly cancellation. In \cite{AC}
the brane match was shown for the case of embedding an $SU(n_1)\times SU(n_2)$
vector bundle in $E_8\times E_8$, resulting in some unbroken gauge group.  
The difference in our situation is that unlike \cite{AC}, for the 
${\cal N}=1$ model of \cite{VW}, at the ${\cal N}=2$ level,
one has to embed an $SU(2)\times SU(2)$ in the $E_8\times E_8$
on the $K3(\times T^2$), a Calabi-Yau 2-fold($\times T^2$), and the resulting $E_7\times E_7$
is Higgsed away, or equivalently, an $E_8\times E_8$ vector bundle is embedded in
$E_8\times E_8$ on the Calabi-Yau 2-fold and not a Calabi-Yau 3-fold. 
The ${\cal N}=1$ Heterotic model on the Voisin-Borcea Calabi-Yau
3-fold with Hodge numbers 11,11 is then obtained  by a suitable ${\bf Z}_2$
modding of the ${\cal N}=2$ model. Hence, it is not that one gets an ${\cal N}=1$ model
by embedding a gauge bundle on a Calabi-Yau 3-fold $Z$ (that may or may not be followed by
a Higgsing), but one gets the required ${\cal N}=1$ model as a three-step process: 
embedding a gauge bundle on a Calabi-Yau 2-fold times $T^2$, 
followed by complete Higgsing away of the resultant gauge group (the embedding and 
Higgsing can be combined into a single step of suitable embedding as discussed above), 
and finally modding by a (freely acting) involution yielding a Calabi-Yau 3-fold $Z$. 
We can still write that the total number of Heterotic moduli is given by the expression:
\begin{equation}
\label{eq:hetmood}
N_{het}=h^{1,1}(Z)+h^{2,1}(Z)+n_{bundle},
\end{equation}
where the bundle moduli correspond to  an involution $\tau$
which acts trivially on the base and as reflection of the fiber
(that can always be defined on an elliptically fibered $Z$ \cite{FMW}).
It no longer can be defined as $h^1(Z,ad(V))=I+2n_o$, where the character-valued index
$I$ is given by -$\sum_{i=}^3(-)^iTr_{H^i(Z,Ad(V))}({1+\tau\over2})=
-\sum_{i=0}^3(-)^ih^i_e(Z,Ad(V))=n_e-n_o$ for no unbroken ${\cal N}=1$ gauge group,
and $e,o$ referring to even,odd respectively under the involution $\tau$. However, given
that such an involution $\tau$ exists, one can still write that
\begin{equation}
\label{eq:gaugebundle}
n_{bundle}=n_e+n_o={\cal I}+2n_o,
\end{equation}
for a suitable ``index" ${\cal I}$. We assume that at the $\tau$-invariant point,
the action of $\tau$ can be lifted to an action of the gauge bundle embedded at
the level of $K3$. This index will encode the information about
$I(K3,Ad(SU(2)\times SU(2)))$ and the Higgsing away of the $E_7\times E_7$, or
equivalently $I(K3,Ad(E_8\times E_8))$ at the
${\cal N}=2$ level, and the freely acting Enriques involution times reflection of $T^2$.
\footnote{As A.Klemm pointed out to us, 
in general, one can always write the index ${\cal I}$ as $a+b\int_{\cal E}c_1^2({\cal E})
+c\int_{\cal E}c_2({\cal E})+d\int_{\cal E}c_1^2({\cal T})+e\int_{\cal E}c_2({\cal T})
+f\int_{\cal E}c_1({\cal E})\wedge c_1({\cal T})$, where $a,b,c,d,e,f$ are constants
and ${\cal T}$ is a line bundle over ${\cal E}$.}
There are no non-perturbative Heterotic 5-branes in the ${\cal N}=1$ model of \cite{VW}.
Hence, for the ${\cal N}=1$ Heterotic/F-theory duality to hold, there will no F-theory
3-branes(given by ${\chi(X_4)\over24}$) either, which implies that 
the elliptically fibered Calabi-Yau 4-fold must 
satisfy the constraint:
\begin{equation}
\label{eq:ENX_4}
\chi(X_4)=0.
\end{equation}
Now,
\begin{eqnarray}
\label{eq:CY1111}
& & h^{1,1}\biggl(CY_3(11,11)\biggr)=11-\int_{{\cal E}}c_1^2({\cal E})=11,
\nonumber\\
& & h^{2,1}\biggl(CY_3(11,11)\biggr)=11+29\int_{{\cal E}}c_1^2({\cal E})=11,
\end{eqnarray}
as $c_1^2({\cal E})=0$. Further, $\int_{{\cal E}} c_2({\cal E})=12$. 
Assuming only a single section of the elliptic fibration: 
$Z\rightarrow_{T^2}{\cal E}(\equiv$ Enriques surface) and no 4-flux,
from general considerations (See \cite{AC}), the Hodge data of $X_4$ will be given by:
\begin{eqnarray}
\label{eq:HodgeX4}
& & h^{1,1}(X_4)=h^{1,1}(Z)+1+r=12-\int_{{\cal E}}c_1^2({\cal E}) + r,\nonumber\\
& & h^{2,1}(X_4)=n_o,\nonumber\\
& & h^{3,1}(X_4)=h^{2,1}+{\cal I}+n_o+1=12+29\int_{{\cal E}}c_1^2({\cal E})+{\cal I}
+h^{2,1}(X_4).
\end{eqnarray}
Now $t\equiv c_1({\cal T})$ (${\cal T}$ being 
a line bundle over $B_2$), the analog of $n$ in the Hirzebruch surface $F_n$, 
is a measure of the non-triviality of the ${\bf CP}^1$-fibration of the rationally ruled $B_3$.
Now, the $CY_3(3,243)$ on the type IIA side, 
can be represented as elliptic fibration over the
Hirzebruch surface $F_n$, where $n$ denotes the non-triviality of fibration
of ${\bf CP}^1_f$ over ${\bf CP}^1_b$. The Weierstrass equation for $n=0$
is given by:
\begin{equation}
\label{eq:F0}
y^2=x^3+\sum_{i=0}^8f^{(8)}_i(z_1)z_2^ix+\sum_{i=0}^{12}g^{(12)}_i(z_1)z_2^i,
\end{equation}
implying that the number of complex structure deformations, $h^{2,1}$ is
given by $9\times9+13\times13-(3+3+1)=243$. 
\footnote{Interestingly, for $n=2$, the
Weierstrass equation is given by:
\begin{equation}
\label{eq:F2}
y^2=x^3+\sum_{i=-4}^4f_{8-4i}(z_1)z_2^{4-i}x+\sum_{i=-8}^8g_{12-2i}(z_1)
z_2^{8-i},
\end{equation}
implying that the number of complex structure deformations, $h^{2,1}$ is
given by $(17+15+13+...+3+1=)81+(25+23+...+3+1=)169-(3+3+1)=243$. Hence,
elliptic fibrations over both $F_0$ and $F_2$ give the same hodge numbers.
We will work with $F_0$.}
Hence, analogous to setting $n=0$, we can  set $t=0$ and doing so would
imply the triviality of the 
fibration: $B_3={\bf CP}^1\times B_2={\bf CP}^1\times{\cal E}$,
for which $h^{2,1}(B_3)=0$ thereby satisfying (\ref{eq:consts1}).

Equating $n_{het}$ to 246, one gets from (\ref{eq:hetmood}) and (\ref{eq:gaugebundle})
the following
\begin{equation}
\label{eq:Iplus2n_o}
{\cal I}+2n_o=224.
\end{equation}
There are no vector multiplets, and in addition to the
heterotic dilaton, $n_{het}$ has to correspond to the number of ${\cal N}=1$
chiral multiplets $n_C$ on the F-theory side. Given that $r=\chi(X_4)=0$, from 
(\ref{eq:nC}) one gets:
\begin{equation}
\label{eq:h21X_4}
h^{2,1}(X_4)=128=n_o.
\end{equation}
This using (\ref{eq:Iplus2n_o}), one gets
\begin{equation}
\label{eq:index}
{\cal I}=-32.
\end{equation}
Using the relation from \cite{SVW}: 
\begin{equation}
\label{eq:constSVW}
{\chi(X_4)\over6}=8+h^{1,1}(X_4)-h^{2,1}(X_4)+
h^{3,1}(X_4),
\end{equation}
one sees that the elliptically 4-manifold $X_4$ that we are
looking for is characterized by:
\begin{eqnarray}
\label{eq:X_4hodges}
& & h^{1,1}(X_4)=12,\nonumber\\
& & h^{2,1}(X_4)=128,\nonumber\\
& & h^{3,1}(X_4)=108.
\end{eqnarray}
This is consistent with (\ref{eq:HodgeX4}).
The $h^{2,2}(X_4)$ can be determined by the following relation
\cite{Klemmetal}
\begin{equation}
\label{eq:h22CY4}
h^{2,2}(X_4)=2(22+2h^{1,1}(X_4)+2h^{3,1}(X_4)-h^{2,1}(X_4))=268,
\end{equation}
which has been obtained from the definitions of elliptic
genera in terms of hodge numbers and as integrals involving
suitable powers of suitable Chern classes, and $c_1(X_4)=0$. 
Hence, ${\cal N}=1$ Heterotic Theory on ${K3\times T^2\over{\bf Z_2}}$ is dual to
F-theory on an elliptically fibered Calabi-Yau 
4-fold: $X_4[h^{1,1}=12,h^{2,1}=128,h^{3,1}=108;0]\rightarrow_{T^2}
{\bf CP}^1\times{\cal E}$. We now discuss an apparent puzzle.
At the ${\cal N}=2$ level,  Heterotic on $K3\times T^2$ should be
dual  to F-theory on $CY_3(3,243)\times T^2$ as a consequence of repeated 
fiberwise application of duality to the basic duality that Heterotic on $T^2$ is dual to
F-theory on $K3$, as well as because type IIA on a $CY_3$ should
be  dual to F-theory on $CY_3\times T^2$ and Heterotic  on $K3\times T^2$ 
is dual to type IIA on  $CY_3(3,243)$. Hence, it is possible that an orbifold 
of $K3\times T^2$ on the Heterotic side should correspond to a suitable orbifold of 
$CY_3\times T^2$ on the F-theory side. Note, however, even though a naive freely acting
orbifold of $CY_3(3,243)\times T^2$ gives the right null Euler Characteristic,
it can not, for instance, give $h^{1,1}=12$, i.e., an enhancement over the
$h^{1,1}(CY_3(3,243)\times T^2)=3+1=4$. This is unlike the case of the F-theory
dual of Heterotic on Voisin-Borcea $CY_3(19,19)$ which corresponded to an involution
with fixed points, considered in \cite{CL}. Of course, given the string/M/F dualities,
the $X_4$ with the derived fibration structure and Hodge data must exist, as 
the F-theory dual corresponding to Heterotic on $CY_3(11,11)$ must exist. One needs to
look further into this issue.
 
The $CY_4$ with the required fibration structure and Hodge data given in 
(\ref{eq:X_4hodges}) and (\ref{eq:h22CY4}) is missing from 
the list of hypersurfaces in ${\bf WCP}^5$ of Kreuzer and Skarke
because it is not possible to get the desired $CY_4$ as a 
hypersurface in any toric variety as fibrations of toric hypersurfaces
have bases that are toric varieties, and the Enriques surface, ${\cal E}$,
is not a toric variety. Perhaps, one needs a ``nef partition" 
(one could use ``nef.x" part of the package PALP\cite{KS}) 
that makes the base, ${\bf CP}^1\times{\cal E}$ a toric hypersurface. One
might have to work with complete intersections in toric varieties.\footnote{We are
grateful to H.Skarke and M.Kreuzer for clarifications on this issue.}

\section{Conclusion}

In this paper, we relate the ${\cal N}=1$ Heterotic theory on a self-mirror 
$CY_3$ to the nonperturbative formulations of type IIA and IIB, namely 
M and F theories. While on the M-theory side, the suitable manifold turned out to 
one of $SU(3)\times {\bf Z}_2$ holonomy, referred to as a `barely $G_2$ manifold',
the elliptically fibered Calabi-Yau 4-fold involves a trivial ${\bf CP}^1$-fibration
over the Enriques surface for its base, and surprisingly has a Hodge data that can
not be obtained as a free involution of (${\cal N}=2$ F-theory on) $CY_3(3,243)\times T^2$.
The orientation-reversing antiholomorphic involution used, both in constructing 
$CY_3$ orientifold on the type IIA side, as well as the barely $G_2$ manifold,
has some interesting properties that become manifest in the large ${\bf CP}^1$-base
limit of $CY_3(3,243)$ which we also discuss.

The precise construction of the $CY_4$ used in the F-theory dual 
and its connection with the ${\cal N}=2$ parent model of F-theory on $CY_3(3,243)\times T^2$, 
needs to be understood.  It will also be interesting to calculate and compare quantities like 
the superpotentials on all sides. 

\section*{Acknowledgements}

We would like to thank S.Govindrajan for bringing \cite{PP}
to our attention, and H.Skarke, M.Kreuzer, A.Grassi, D.Joyce, H.Partouche, E.Scheidegger
 and especially
S.Kachru for useful communications and clarifications. We are particularly grateful to
A.Klemm for several very useful discussions. We thank D.L\"{u}st for going through
a preliminary draft of the manuscript and making useful comments, and S.Bhattacharyya
for pointing out the typos in the earlier version. 
This research was supported by the Alexander von Humboldt foundation.

\end{document}